\documentclass[showpacs, pra,twocolumn,preprintnumbers ,amsmath, amssymb, superscriptaddress, aps]{revtex4-2}
\usepackage{color}
\usepackage{amsmath,amssymb}
\usepackage{pifont}
\usepackage{amssymb}  
\usepackage{bbold}
\usepackage{float}
\usepackage{subfloat}

\usepackage[caption=false]{subfig}
\usepackage{tikz}
\usepackage{makecell}
\usepackage{subfig}
\usepackage{pifont}   
\usepackage{graphicx} 
\graphicspath{{Figures2/}}
\usepackage{dcolumn}  
\usepackage{bm}       
\usepackage{multirow} 
\usepackage{placeins}
\usepackage[colorlinks]{hyperref}
\usepackage{mathtools}
\usepackage{appendix}

\def \be{\begin{align}}
	\def \ee{\end{align}}
\def \bea{\begin{eqnarray}}
	\def \eea{\end{eqnarray}}

\begin{document}
	
	\title{Electron trapping via magnetic and laser fields in  gapped graphene quantum dots}
	
	\author{Ahmed Bouhlal}
	\affiliation{Laboratory of Theoretical Physics, Faculty of Sciences, Choua\"ib Doukkali University, PO Box 20, 24000 El Jadida, Morocco}

	\author{Mohammed El Azar}
	\affiliation{Laboratory of Theoretical Physics, Faculty of Sciences, Choua\"ib Doukkali University, PO Box 20, 24000 El Jadida, Morocco}
	
	\author{Aotmane En Naciri} 
	\affiliation{LCP-A2MC, Université de Lorraine, ICPM, 1 Bd Arago 57070 Metz, France}
	
	\author{Elmustapha Feddi}
	\affiliation{School of Applied and Engineering Physics, Mohammed VI Polytechnic University, Lot 660, Hay Moulay Rachid, 43150,Morocco, Ben Guerir}
	
		\author{Ahmed Jellal}
	\affiliation{Laboratory of Theoretical Physics, Faculty of Sciences, Choua\"ib Doukkali University, PO Box 20, 24000 El Jadida, Morocco}

	\begin{abstract}
We study electron scattering in graphene quantum dots (GQDs) under the combined influence of a magnetic field, an energy gap, and circularly polarized laser irradiation. Using the Floquet approach and the Dirac equation, we derive the energy spectrum solutions. The scattering coefficients are calculated explicitly by matching the eigenspinors at the GQD interfaces, revealing a dependence on several physical parameters. In addition, we compute the scattering efficiency, the electron density distribution, and the lifetime of the quasi-bound states.
Our numerical results show that the presence of an energy gap and circularly polarized laser irradiation enhances the localization of the electron density within the GQDs, leading to an increase in the lifetime of the quasi-bound states. In particular, the intensity and polarization of the light influence the scattering process, allowing the manipulation of the electron confinement state. These results highlight the importance of combining magnetic fields and polarized light to control electronic transport in graphene nanostructures.	 
	 

	\end{abstract}

	\maketitle
	
	\section{Introduction}
A unique two-dimensional material, graphene exhibits a distinctive hexagonal lattice structure in which carbon atoms are interconnected by covalent bonds \cite{Novoselov04,Geim09,Allen10}. What distinguishes this material is its extraordinary energy spectrum: at low energy levels, charged particles exhibit behavior characteristic of massless Dirac fermions \cite{Castro10,Zhang05,Katsnelson07}. This remarkable property provides an unprecedented opportunity to study relativistic phenomena using experiments with electrons at energy levels traditionally associated with non-relativistic physics \cite{novoselov2006unconven,Beenakker08,das2011electronic}.
After the first synthesis of graphene, researchers have extensively studied its interactions with external fields, leading to remarkable discoveries. This extraordinary material has proven to be an ideal platform for exploring fundamental physical phenomena. Studies have revealed numerous quantum effects, including Landau quantization \cite{Guinea06,Zhang22}, the intricate Hofstadter butterfly spectrum \cite{Kooi18,Lian21,Benlakhouy22}, Klein tunneling \cite{Katsnelson06}, and quantum Hall behavior \cite{Zhang05,Yin22}. In addition, researchers have demonstrated the potential of graphene to study quantum interference through Aharonov-Bohm effects \cite{Recher07,AzarAB24}. Beyond fundamental physics, graphene has shown exceptional promise for technological applications, particularly in electronics and optoelectronics. These applications include devices ranging from field-effect transistors and phototransistors to advanced light detection and optoelectronic systems \cite{Lin23,Wu23,Jiang22}.

Since the discovery of graphene, there has been considerable research interest in electrostatic quantum dots \cite{Zhangrecent22,Cao23}. However, the implementation of electrically driven QDs poses significant challenges due to the Klein tunneling effect, which complicates the application of local gates commonly used in traditional two-dimensional electron systems \cite{Chen23,Wu23}. A promising alternative approach involves the use of magnetic fields for electron confinement \cite{Wu23ultrafast,ai2022high}. Recent research has demonstrated the existence of localized states with discrete energy spectra in magnetically confined QDs, providing an elegant solution to circumvent the Klein tunneling problem \cite{zhang2022electron,anwar2021coherent}. 
Recent research has demonstrated the existence of localized states with discrete energy spectra in magnetically confined QDs, providing an elegant solution to circumvent the Klein tunneling problem \cite{zhou2021electronic,Penadriven22}. Magnetic confinement can be achieved by locally modulating a uniform magnetic field \cite{shang2023light}, a technique that has been extensively studied both theoretically and experimentally \cite{Zarenia11,Grujic2011,Orozco19,Belokda23}. The confinement of electrons in graphene QDs remains a particularly active area of research \cite{Grushevskaya21,Azar24life}, driven by its crucial importance for future technological applications. Although Klein tunneling still poses a challenge for trapping electrons at normal incidence on graphene QDs, recent advances have demonstrated the possibility of transient trapping \cite{Penalight,wang2022fullerene}, opening new avenues for optimizing these confinement times.

Our study investigates the enhancement of electron confinement in graphene quantum dots (GQDs) under the combined influence of a uniform magnetic field, an energy gap, and laser irradiation. A theoretical framework is developed using the Dirac-Weyl formalism, which captures the unique behavior of massless Dirac fermions in graphene through the minimal coupling approach \cite{Goerbig11}. Building on previous investigations \cite{Azar24life} of Dirac fermions in magnetic field, we extend the analysis to include the effects of circularly polarized light on the quasi-bound states within the GQD. This approach allows us to study how the interplay between magnetic confinement and laser-induced effects modifies the electron dynamics and scattering properties in these quantum structures.
Our study explores the enhancement of electron confinement in graphene quantum dots (GQDs) under the combined effects of a uniform magnetic field, an energy gap, and laser irradiation. We develop a theoretical framework based on the Dirac-Weyl formalism, which effectively describes the unique behavior of massless Dirac fermions in graphene through the minimal coupling approach \cite{Goerbig11}. Extending previous studies \cite{Azar24life} of Dirac fermions in magnetic fields, we extend the analysis to include the effect of circularly polarized light on the quasi-bound states within the GQD. This allows us to study how the interaction between magnetic confinement and laser-induced effects alters the electron dynamics and scattering properties in these quantum structures. 
In addition, the complex energy approach facilitates the quantitative assessment of electron trapping times, providing critical insight into the temporal stability of confined states. The analysis explores the interplay between several key physical parameters, including incident electron energy, quantum dot radius, magnetic field strength, energy gap, light intensity, and light polarization. Our theoretical framework allows us to study how these parameters collectively influence the electron confinement and scattering processes. The results show the emergence of enhanced scattering resonances with increasing energy gap, while the combined effects of magnetic field and laser parameters create favorable conditions for strong electron localization. Through a mode-specific analysis, we investigate how different angular momentum channels respond to these control parameters, providing deeper insights into the quantum mechanical nature of electron confinement in these structures. 

The present paper is organized as follows. In Sec. \ref{theory}, a theoretical model that accurately describes the electron scattering process is presented. The solutions of the Dirac equation are obtained to determine the eigenstates in the different regions of the system, considering a confinement potential, the influence of a uniform magnetic field, and the presence of a circularly polarized laser wave. In Sec. \ref{SSPP}, the continuity conditions at the interfaces are used to obtain analytical expressions for the key physical quantities of the scattering process. These include scattering coefficients, efficiencies, electron density distributions, and quasi-bound state lifetimes. Sec. \ref{res} is devoted to the numerical analysis of the theoretical results, exploring the influence of various physical parameters characterizing the system, such as the incident energy, the quantum dot radius, the magnetic field strength, the energy gap, the light polarization, and the light intensity. Finally, Sec. \ref{cc} provides a comprehensive summary of the main results and conclusions, emphasizing the impact of the energy gap and light polarizations on the quasi-bound states and electron localization in the graphene quantum dot.

%

	\section{Hamiltonian formalism}\label{theory}
	
\begin{figure}[ht]
		\centering
		\includegraphics[scale=0.3]{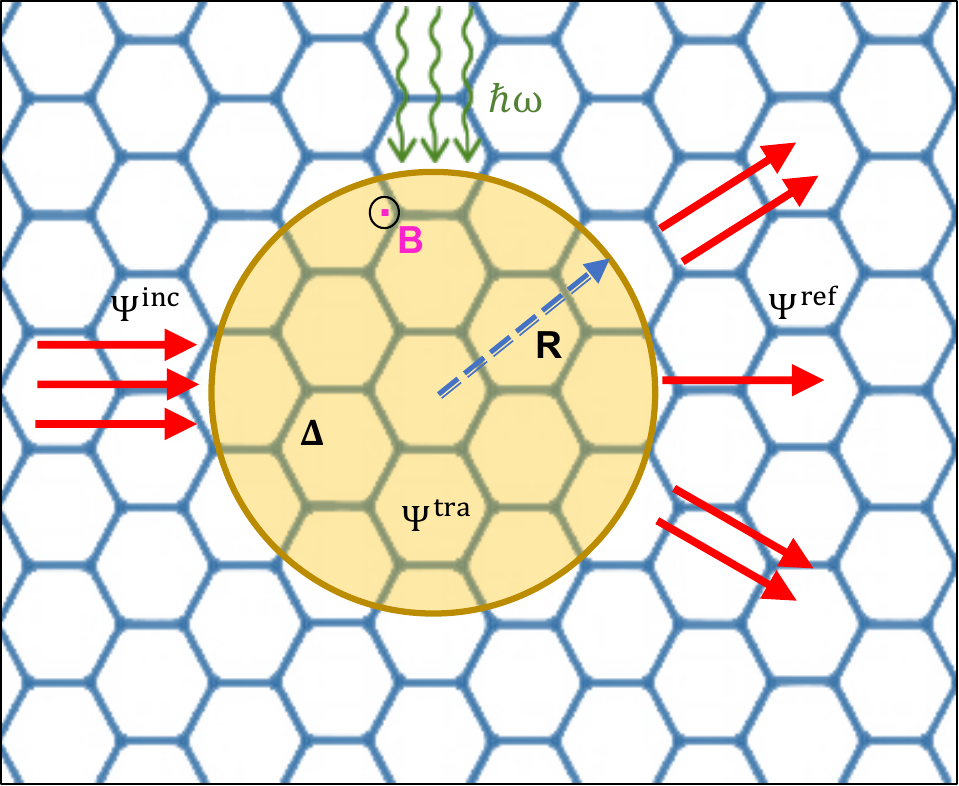}
		\caption{The profile consists of a graphene quantum dot of radius \( R \) subjected to an energy gap \( \Delta \), a magnetic field \( B \), and a laser field of frequency \( \omega \).}\label{fig1}
		\end{figure}

	We consider a system of graphene quantum dots (GQDs) exposed to both magnetic and laser fields, along with an energy gap as depicted in Fig. \ref {fig1}.
	The single-valley Hamiltonian governing the motion of the charge carriers in the GQDs with radius \( R \) can be expressed as follows
	\begin{equation}\label{Hamilt}
		H =v_F \vec \sigma \cdot \left(\vec p +e \left[\vec A + \vec A(t)\right]\right)+ \Delta \sigma_z
	\end{equation}
	where $v_F = 10^6$ ms$^{-1}$  is the Fermi velocity, and $\vec \sigma=(\sigma_x ,\sigma_y. \sigma_z) $ are  Pauli matrices, and $\Delta$ is an external energy gap that can be created by a substrate such as Boron Nitride.
	The vector potential ${\vec{A}}(t)$ of the laser field in the dipole approximation \cite{Loudon2000}
	is generated by an electric field $E(t)$ of amplitude $F$ and frequency $\omega$ in arbitrarily polarization, which is given  by
%
	\begin{align}
		{\vec{A}}(t)=A_0\left(\cos(\omega t)+\varrho 
		\cos\varphi(t), -\sin(\omega t)+\varrho \sin\varphi(t)\right)
	\end{align}
	and amplitude $A_0$ is linked to $F$, with $\varphi(t)=\omega t+\theta$.
	It convenient to select the vector potential $\vec A
 = \frac{B}{2} (x, y)$ in symmetric gauge. 
 The Floquet formalism is used because of ${\vec{A}}(t) $ associated with the laser field causes the Hamiltonian \eqref{Hamilt} to exhibit time periodicity \cite{shirley1965,li2018,wurl2018,giovannini2020,junk2020}. The associated regime becomes non-resonant when the applied light energy exceeds the Dirac fermion energy scale. Consequently, a stationary effective Floquet Hamiltonian is obtained, which is given by
 \begin{align}
 &	H_\text{eff}=v_F \vec \sigma \cdot \left(\vec p +e \vec A(\vec r)\right)+ \Delta \sigma_z+\frac{1}{\hbar\omega}
 	[H_{-1},H_{1}]\label{ham}\\
 &	H_n=ev_F\frac{1}{T}\int_{0}^{T}e^{in\omega 
 		t}
 	\vec{\sigma}\cdot{\vec{A}}(t)	
 		dt
 \end{align}
 where $T=\frac{2\pi}{\omega}$ is  a period  of the light, $H_{-1}$ and $H_{1}$ describe the virtual process of process of absorption and emission of a photon. They satisfy the commutation relation
 \begin{equation}
 	[H_{-1},H_{1}]=-(ev_FA_0)^2(\varrho^2-1)\sigma_z.
 \end{equation} 
 As a result, the Hamiltonian \eqref{Hamilt}  can be transformed into
\begin{equation}\label{Hamilt2}
		H_\text{eff}=v_F \vec \sigma \cdot \left(\vec p +e \vec A\right)+ \left[\Delta - \frac{(ev_FA_0)^2}{\hbar\omega}(\varrho^2-1)\right] \sigma_z.
\end{equation}	
 Because of the spherical symmetry, we can write the Hamiltonian \eqref{Hamilt2} in polar coordinates $(r,\theta)$. To proceed, we first introduce the matrices 
	\begin{equation}
		\sigma_r = \begin{pmatrix} 0 & e^{-i\theta}\\  e^{i\theta} &0 \\
		\end{pmatrix}, \quad 
		\sigma_\theta = 
		\begin{pmatrix} 0 & -ie^{-i\theta}\\  ie^{i\theta} &0 \\
		\end{pmatrix}
	\end{equation}
	and therefore we get
	\begin{widetext}
	\begin{equation}\label{Hamilt3}
		H_\text{eff} = \begin{pmatrix} \Gamma & -i\hbar v_F  e^{-i\theta}\left(\partial_r - \frac{i}{r}\partial_\theta - \frac{e Br}{2\hbar}\right)\\
			-i\hbar v_F  e^{-i\theta}\left(\partial_r + \frac{i}{r}\partial_\theta + \frac{e Br}{2\hbar}	\right)&-\Gamma \\
		\end{pmatrix}
	\end{equation}	
\end{widetext}
	where we have set an new energy gap 
	\begin{equation}
		\Gamma= \Delta - \frac{(ev_FA_0)^2}{\hbar\omega}(\varrho^2-1).
	\end{equation}

	Since the total angular momentum operator $J _z= -i\hbar \partial_\theta + \frac{\hbar}{2}\sigma_z$ commutes with the Hamiltonian \eqref{Hamilt3}, then  the eigenspinors can be expressed as
	\begin{equation}\label{ansatz}
		\Psi(r,\theta)= \dbinom{e^{im\theta}\psi^+(r) }{ie^{i(m+1)\theta}\psi^-(r)}  
	\end{equation}
where the quantum numbers $m$ are the eigenvalues of the operator $J_z$. To proceed further, we use  the eigenvalue equation $H_\text{eff} \Psi(r,\theta) = E \Psi(r,\theta)$ to get
		\begin{align}
			&			\left(\partial_r + \frac{r}{2 l_B^2} -\frac{m}{r}\right) \psi^+(r) =-\frac{E+\Gamma}{\hbar v_F} \psi^-(r) \label{8a}\\
			&	\left( \partial_r+\frac{m+1}{r} -\frac{r}{2l_B^2}\right) \psi^-(r) = \frac{E-\Gamma}{\hbar v_F} \psi^+(r) \label{8b}.
		\end{align}
	By substituting \eqref{8a} into \eqref{8b}, we end up with a second order differential equation for $\psi^+(r)$
	\begin{equation}\label{e:9}
		\left( \partial_r^2+\frac{1}{r}\partial_r  + \frac{m+1}{l_B^2}-\frac{r^2}{4 l_B^2}-\frac{m^2}{r^2}+\kappa^2\right)\psi^+(r) =0
	\end{equation}
	where the wave vector depending on energy gap $\Delta$ and laser parameters $(A_0,\omega) $ is
	\begin{align}
	\kappa=\frac{\sqrt{\vert E^2-\Gamma^2\vert}}{\hbar v_F}
	\end{align}
	and   $l_B= \sqrt{\frac{\hbar}{eB}}$ is the magnetic length.

	To solve \eqref{e:9}, we begin by analyzing the asymptotic limits that determine the required physical behavior as a function of $r$.
	Indeed, for $r\rightarrow \infty$,  \eqref{e:9} reduces to
	\begin{equation}\label{e:10}
		\left( \partial_r^2+\frac{1}{r}\partial_r - \frac{r^2}{4 l_B^2}\right) \psi^+(r) =0
	\end{equation}
which corresponds to the modified Bessel equation for the zero-order case, and thus the solution is given by
	\begin{equation}
	\psi^+(r)  =c_1 I_0\left( \frac{r^2}{4l_B^2}\right) +c_2 K_0\left( \frac{r^2}{4l_B^2}\right) 
	\end{equation}
where $I_0(x)$ and $K_0(x)$ represent the zero-order modified Bessel functions of the first and second kinds, respectively. We set $c_1 = 0$ and $c_2 = 1$ to prevent the divergence of $I_0(x)$ as $x$ approaches infinity. Now, applying the asymptotic behavior $K_0(x) \underset{x\gg 1}{\sim} \frac{e^{-x}}{\sqrt{x}}$, we can approximate $\psi^+(r)$ as
	\begin{equation}\label{e:13}
	\psi^+(r)	 \sim 2 l_B\frac{e^{-\frac{r^2}{4 l_B^2}}}{r}.
	\end{equation}
	In the opposite limit, as $r \rightarrow 0$, we derive the following expression from \eqref{e:9}
	\begin{equation}\label{e:14}
		\left(  \partial_r^2+\frac{1}{r}\partial_r  - \frac{m^2}{r^2} \right) \psi^+(r)=0
	\end{equation}	
and   solution	is given by
	\begin{equation}\label{e:15}
	\psi^+(r) =\frac{c_3}{2}(r^m +r^{-m})+\frac{i c_4}{2}(r^m -r^{-m})
	\end{equation}	
where $c_3$ and $c_4$ must be chosen to ensure that the solution satisfies the physical constraints. Therefore, we analyze the cases for positive and negative values of $m$ separately. For $m \ge 0$, the term $\sim r^{-m}$ must vanish, resulting in $c_4 = -ic_3$, by convention, we set $c_3 = \frac{1}{2l_B^{\pm m}}$. For $m < 0$, the term $\sim r^m$ must vanish, then we set $c_4 = ic_3$ and again choose $c_3 = \frac{1}{2l_B^{\pm m}}$. Combining these conditions, we express the asymptotic behavior of $\psi^+(r)$ as
		\begin{align} \label{16a}
		\psi^+(r)=	\begin{cases}	
			 \left(\frac{r}{2l_B}\right)^m	,& m\ge 0\\ 
			\left(\frac{r}{2l_B}\right)^{ -m},& m<0. 
			\end{cases}
		\end{align}
The asymptotic behaviors of \eqref{e:9} suggest expressing the solution in the following form
	\begin{equation}\label{e:19}
	\psi^+_{\pm} (r) =\left(\frac{r}{2l_B}\right)^{\pm m }  \frac{e^{-r^2/4 l_B^2}}{r/2 l_B} \zeta_{\kappa\pm}^{+}(r)
	\end{equation}
	where the $\pm$ sign corresponds to $m \ge 0$ and $m < 0$, respectively.
	By performing the variable change $ \xi = \frac{r^2}{2 l_B^2} $ and applying the transformation $ \zeta_{\kappa\pm}^{+}(\xi) = \sqrt{\xi} \chi_{\kappa\pm}^{+}(\xi)$, we map \eqref{e:19} as 
	\begin{equation}\label{e:20}
	\psi^+_{\pm} (\xi)	 = 
	\xi^{\pm m/2 }  e^{-\xi/2} \chi_{\kappa\pm}^{+}(\xi)
	\end{equation}
	and then we have to determine the function $\chi_{\kappa\pm}^{+}(\xi)$ to finally get the solution. In fact, by returning to \eqref{e:9} and doing some algebraic manipulations, we arrive at Kummer-type differential equations given by
		\begin{align}
			&\left[\xi \partial_\xi^2  +\left( m+1-\xi\right)  \partial_\xi  +\frac{l_B^2 \kappa^2}{2}\right] \chi_{\kappa+}^{+}(\xi)=0  \label{21a}\\
			&\left[\xi \partial_\xi^2  +\left( 1-m-\xi\right)  \partial_\xi 
			+ m+\frac{l_B^2 \kappa^2}{2} \right] \chi_{\kappa-}^{+}(\xi)=0 \label{21b} 
		\end{align}
which lead to the confluent hypergeometric functions as solution
		\begin{align}
			&\chi_{\kappa+}^{+}(\xi)=\prescript{}{1}{F}_1^{}\left(-\frac{l_B^2 \kappa^2}{2},m+1,\xi\right) \label{18a}\\
			&\chi_{\kappa-}^{+}(\xi)=\prescript{}{1}{F}_1^{}\left(-l-\frac{l_B^2 \kappa^2}{2},1-m,\xi\right) \label{18b}.
		\end{align}
	As a result, by combining all the previous findings, we derive the solutions to the second-order differential equation \eqref{e:9}. These solutions are expressed in a separated form as 
		\begin{align}
			&\psi^{+}_+=\left(\frac{r}{2l_B}\right)^{\vert m\vert} e^{-r^2/4 l_B^2}\prescript{}{1}{F}_1^{}\left(-\frac{l_B^2 \kappa^2}{2},m+1,\frac{r^2}{2 l_B^2}\right) \label{22a}\\
			&\psi^{+}_-=\left(\frac{r}{2l_B}\right)^{\vert m\vert} e^{-r^2/4 l_B^2}\prescript{}{1}{F}_1^{}\left(-m-\frac{l_B^2 \kappa^2}{2},1-m,\frac{r^2}{2 l_B^2}\right) \label{22b} 
		\end{align}
where each component corresponds to a specific physical quantity governing the behavior of the system. The separate form allows for a more detailed analysis of the contributions of individual parameters, providing insight into the underlying dynamics and facilitating further exploration of boundary conditions and mode-specific properties. 
	The other components of the spinor \eqref{ansatz} can be derived by inserting \eqref{22a} and \eqref{22b} into \eqref{8a}. This procedure leads to the following expression
				\begin{widetext}
			\begin{align}
				\psi^{-}_+=&\frac{\kappa}{2(m+1)}\left(\frac{r}{2l_B}\right)^{\vert m\vert+1} e^{-r^2/4 l_B^2}\prescript{}{1}{F}_1^{}\left(1-\frac{l_B^2 \kappa^2}{2},m+2,\frac{r^2}{2 l_B^2}\right) \label{24a}\\
				\psi^{-}_-=&\frac{1}{\kappa}
				\left(\frac{r}{2l_B}\right)^{\vert m\vert-1}
				 e^{-r^2/4 l_B^2}\\
				 &\left[2m \prescript{}{1}{F}_1^{}\left(-m-\frac{l_B^2 \kappa^2}{2},1-m,\frac{r^2}{2 l_B^2}\right)+ \frac{( 2m+l_B^2 \kappa^2)r^2} {2( 1-m) l_B^2}\prescript{}{1}{F}_1^{}\left(1-m-\frac{l_B^2 \kappa^2}{2},2-m,\frac{r^2}{2 l_B^2}\right)\right] \notag \label{24b} .
			\end{align}
				\end{widetext}
In the following analysis, we will explore how the above results can be applied to study the scattering phenomenon by considering various quantities.  In addition, we will analyze the effect of laser field on the scattering process and discuss how the asymptotic forms of the wavefunction influence the final result. This approach will allow us to gain deeper insights into the physical properties of the system and its response to external perturbations.
	
	\section{Scattering problem}\label{SSPP}
	
	  To analyze the scattering problem, we will first discuss how an electron scatters from gapped GQDs of radius \( R \) in the presence of magnetic and laser fields. Before identifying the key parameters that characterize the scattering, consider an electron moving in the \( x \) direction with energy \( E = \hbar v_F k \), where \( k \) is the corresponding wave number. The incident electron can thus be described by a plane wave as 
	  \begin{widetext}
	\begin{equation}\label{e:25}
		\Psi_k^{\text{inc}}(r,\theta) =\frac{1}{\sqrt{2}}e^{ikr\cos\theta }\dbinom{1}{1}=\frac{1}{\sqrt{2}}\sum_{m=-\infty}^{\infty}i^m \dbinom{e^{im \theta}J_m (kr) }{ie^{i(m+1)\theta}J_{m+1}(kr)}
	\end{equation}
\end{widetext}
	where \( J_m(z) \) is the Bessel function of the first kind. 
	 Since the reflected electron wave must satisfy the infinite boundary conditions imposed by the scattering mechanism under study, we decompose it into partial waves
	\begin{equation}\label{e:26}
		\Psi_k^{\text{ref}}(r,\theta) =\frac{1}{\sqrt{2}}\sum_{m=-\infty}^{\infty} a_m^r i^m \dbinom{e^{im\theta}H_m (kr) }{ie^{i(m+1)\theta}H_{m+1}(kr)}
	\end{equation}
such that \( H_m(x) \) is the Hankel function of the first kind, which is a linear combination of the Bessel function \( J_m(x) \) and the Neumann function \( Y_m(x) \), namely, \( H_m(x) = J_m(x) + i Y_m(x) \). 
	The transmitted solution can be obtained from the previous analysis as
	\begin{widetext}
	\begin{equation}\label{e:28}
		\Psi_\kappa^{\text{tra}}(r,\theta) =\sum_{m=-\infty}^{-1} a_m^{t-}  \dbinom{e^{im\theta}\psi_{\kappa-}^{+} (r) }{i e^{i(m+1)\theta}\psi_{\kappa-}^{-}(r)} +\sum_{m=0}^{\infty} a_m^{t+ } \dbinom{e^{im\theta}\psi_{\kappa+}^{+} (r) }{ie^{i(m+1)\theta}\psi_{\kappa+}^{-}(r)}
	\end{equation}
	\end{widetext}
	where \( \kappa \) represents the wave number associated with the electron inside the GQDs, as shown in Fig. \ref{fig1}. The coefficients \( a_m^{r} \) and \( a_m^{t} \) will be determined explicitly by applying the boundary condition $r=R$. More specifically, we have
	\begin{equation}\label{e:29}
		\Psi_k^{\text{inc}}(R,\theta) +\Psi_k^{\text{ref}}(R,\theta) =\Psi_\kappa^{\text{tra}}(R,\theta) 
	\end{equation}
and thus, after replacing the spinors, we obtain
		\begin{align}
			&	 i^m J_m(kR)+ i^m a_m^{r\pm} H_m(kR) =\sqrt{2} a_m^{t\pm} \psi_{\kappa\pm}^{+} ( R) \label{30a}\\
			&	 i^{m+1} J_{m+1}(kR)+ i^{m+1} a_m^{r\pm} H_{m+1}(kR) = \sqrt{2}i a_m^{t\pm} \psi_{\kappa\pm}^{-} ( R)  \label{30b} 
		\end{align}
which can be solved to end up with the coefficients
		\begin{align}
			a_m^{t\pm}&=\frac{i \sqrt{2} e^{im\pi/2}}{\pi \kappa R[H_m(kR)\psi_{\kappa\pm}^{-} ( R)-H_{m+1}(kR)\psi_{\kappa\pm}^{+} ( R) ]} \label{31a}\\
			a_m^{r\pm}&=\frac{-J_m(kR)\psi_{\kappa\pm}^{-} ( R)+J_{m
					+1}(kR)\psi_{\kappa\pm}^{+} (\kappa R)}{H_m(kR)\psi_{\kappa\pm}^{-} ( R)-H_{m
					+1}(kR)\psi_{\kappa\pm}^{+} ( R) }  \label{31b} .
		\end{align}

Starting from to the Hamiltonian \eqref{Hamilt}, we show that the current density is  $\vec j= \Phi^\dag \vec \sigma \Phi$ where $\Phi=\Psi_\kappa^{\text{tra}}$ is inside  and $\Phi=\Psi_k^{\text{inc}}+\Psi_k^{\text{ref}}$ is outside the GQDs. Consequently, the radial component  of the current density is
\begin{align}\label{RCCD}
		j_{\text{rad}}^r= \vec j \cdot \vec e_r=
		\left(\Psi_k^{\text{ref}}\right)^{*}(r,\theta)\begin{pmatrix}			0 & e^{-i\theta} \\
			e^{i\theta} & 0
		\end{pmatrix} \Psi_k^{\text{ref}}(r,\theta).
\end{align}	
For later use, we can simplify \( j_{\text{rad}}^r \) by considering large values of \( kr \), i.e., \( kr \gg 1 \). In this limit, we can approximate \( H_m(kr) \) as
	\begin{equation}\label{e:27}
		H_m(kr) {\sim} \sqrt{\frac{2}{\pi kr}} e^{i(kr-\frac{l\pi}{2}-\frac{\pi}{4})}.
	\end{equation}	
As a result, we show that \eqref{RCCD} takes the form
	\begin{align}\label{e:32}
		j_{\text{rad}}^r =&\frac{4}{\pi k r}\sum_{m=-\infty}^{+\infty} \vert a_m^r\vert^2 \\
		&+\frac{8}{\pi k r}\sum_{m<m'} \Re(a_m^r a_{m'}^r)\cos[(m-m')\theta].\nonumber
	\end{align}
    Now we define other quantities to emphasize the basic features of our system. Indeed, in the limit $k r \to \infty$, \eqref{e:32} is used to calculate the effective scattering cross section $\sigma$ is
	\begin{equation}
		\sigma=\frac{I_{\text{rad}}^r}{I^{\text{inc}} / A_u}
	\end{equation}
where the incident flux per unit area is $I^{\text{inc}} / A_u$, and the total reflected flux over the GQDs of radius $R$ is $I_{\text{rad}}^r$. We show that the total reflected flux $I_{\text{rad}}^r$ is given by
	\begin{equation}
		I_{\text{rad}}^r=\int_0^{2 \pi} j_{\text{rad}}^r r d \theta=\frac{8}{k} \sum_{m=-\infty}^{+\infty}\left|a_m\right|^2
	\end{equation}
and  $I^i / A_u=1$ for 
 the incident wave  \eqref{e:25}.
	To enhance our study of the scattering problem of Dirac fermions in circular quantum dots of various sizes, we analyze the scattering efficiency \( Q \). This is defined as the ratio of the scattering cross section to the geometrical cross section
	\begin{equation}\label{e:33}
		Q=\frac{\sigma}{2 R}=\frac{4}{k R}\sum_{m=-\infty}^{+\infty} \vert a_m^r\vert^2 .
	\end{equation}
	Recall that the coefficients \( a_m^r \) in \eqref{31b} depend on the gap \( \Delta \), the magnetic field \( B \), the laser amplitude \( A_0 \), and the polarization \( \varrho \). This dependence provides several configurations of the physical parameters, allowing us to explore and discuss different aspects of the scattering phenomenon occurring in our system.  
	
	\section{RESULTS AND DISCUSSION} \label{res}

We present a comprehensive analysis of electron scattering in the gapped  GQDs subjected to the combined effects of a magnetic field and a laser field. The scattering efficiency \(Q\) \eqref{e:33} is used as the primary parameter to characterize the scattering properties of the system. Our study systematically investigates the influence of several key parameters: quantum dot radius \(R\), magnetic field strength \(B\), energy gap \(\Delta\), light intensity \(I_L\), and light polarization \(\varrho\). Throughout our study, laser irradiation is characterized by its intensity, denoted as \(I_L\), which is directly proportional to the square of the amplitude of the potential vector of the electromagnetic wave. The intensity is  \(I_L = \epsilon_0 \omega^2 A_0^2\), where \(\epsilon_0\) is the vacuum permittivity and \(\omega\) is the angular frequency of the light wave. \(I_L\) serves as a control parameter to examine the effect of laser irradiation on electron transport and localization in the GQDs. In fact, it allows us to evaluate how light modulates electron confinement. In contrast, the light frequency is fixed at \(\omega = 5 \times 10^{14} \, \text{s}^{-1}\), and the incident electron energy is kept at \(E = 20\) meV, in the low-energy regime where the unique properties of graphene are most pronounced. Our numerical results will reveal a variety of scattering phenomena, including distinct resonances and intricate relations with the physical parameters. Special emphasis will be given to the analysis of different excited modes and their influence on  \(Q\).

\subsection{Scattering efficiency}

	\begin{figure}[H]
		\centering
		\includegraphics[scale=0.165]{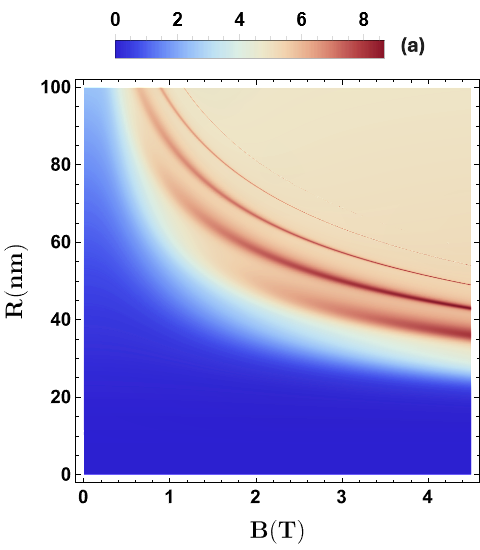}
		\includegraphics[scale=0.165]{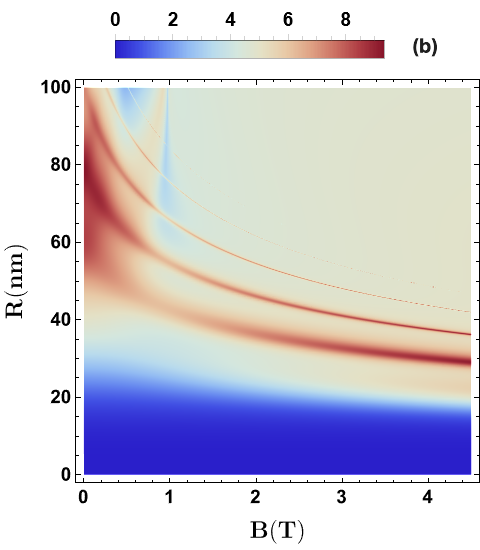}
		\includegraphics[scale=0.165]{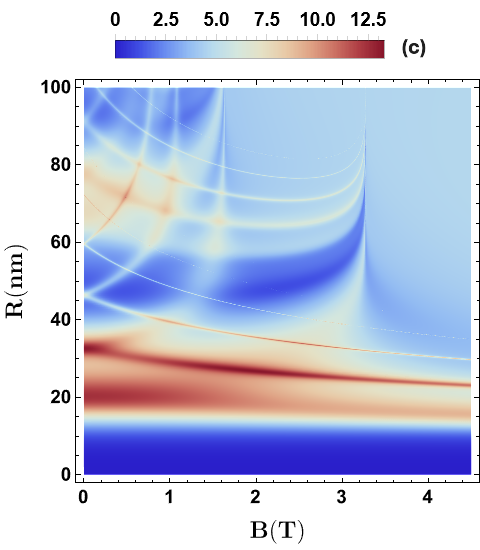}\\
			\includegraphics[scale=0.165]{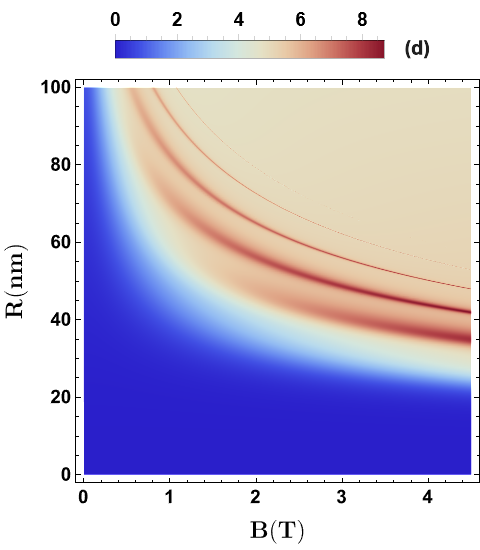}
			\includegraphics[scale=0.165]{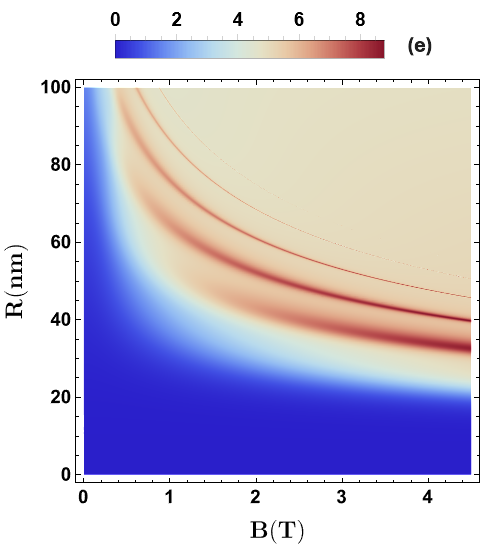}
			\includegraphics[scale=0.165]{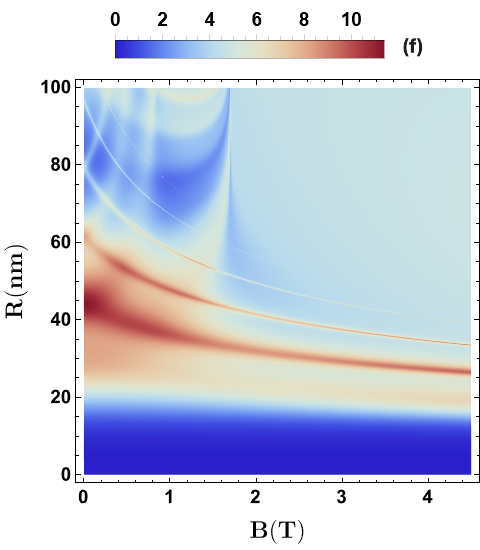}\\
				\includegraphics[scale=0.165]{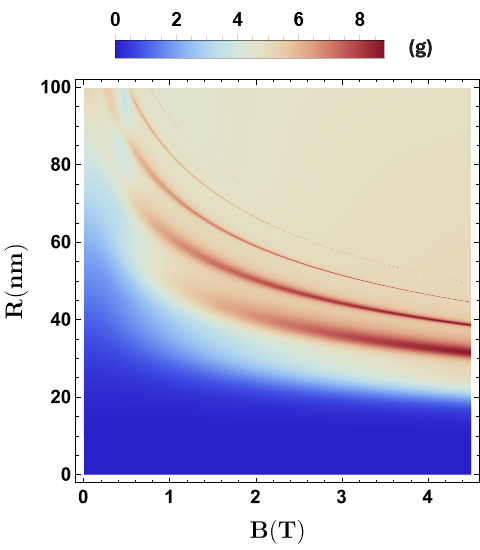}
				\includegraphics[scale=0.165]{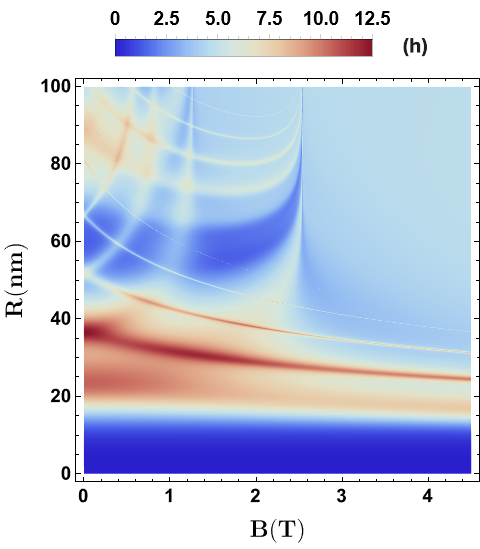}
				\includegraphics[scale=0.16]{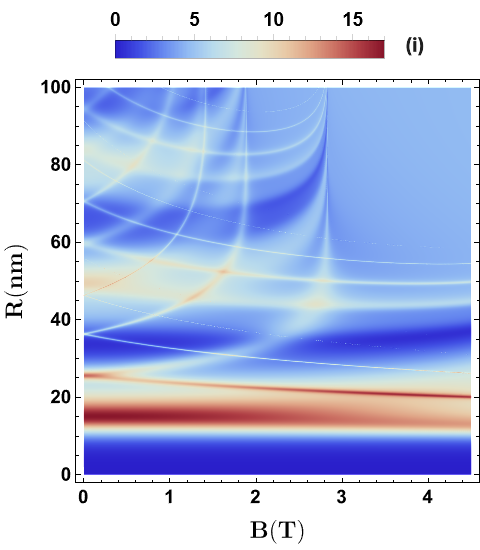}\\
					\includegraphics[scale=0.165]{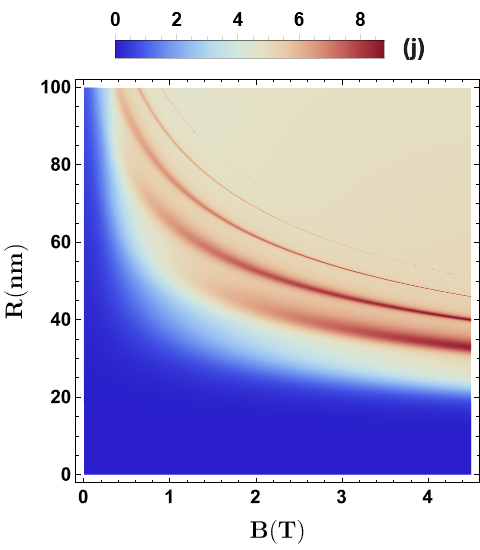}
					\includegraphics[scale=0.165]{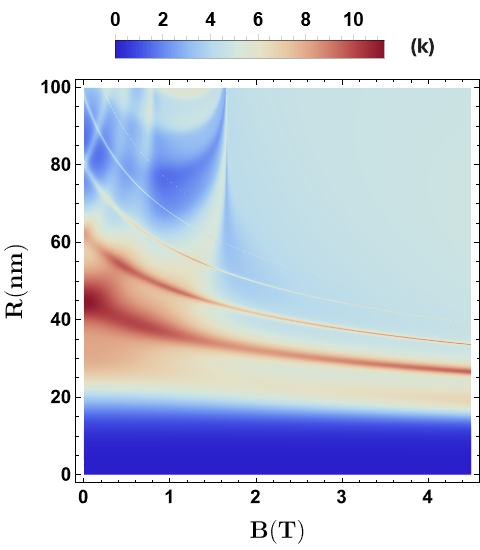}
					\includegraphics[scale=0.165]{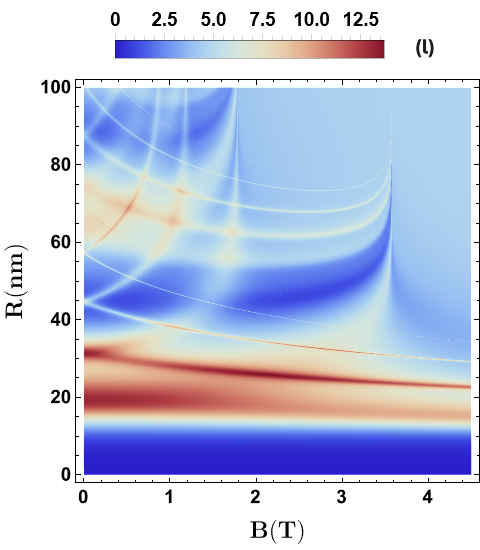}
		\caption{The scattering efficiency $Q$ as a function of the radius $R$ and magnetic field strength $B$ for the incident energy $E = 20$ meV, energy gap [(a,b,c,d,e,f): $\Delta = 0$, (g,h,i,j,k,l): $\Delta = 20$ meV], light polarization
			[(a,b,c,g,h,i): $\varrho = 0$, (d,e, f, j, k, l): $\varrho = 0.5$], and light intensity  [(a,d,g,j): $I_L= 1$ W/cm², (b,e,h,k):  $I_L = 3$ W/cm², (c,f,i,l): $I_L = 5$ W/cm²].}
		\label{fig2}
	\end{figure}
	
	Fig. \ref{fig2} shows density plots of the scattering efficiency \( Q \) as a function of radius \( R \) and magnetic field strength \( B \) for an incident electron energy of \( E = 20 \) meV, with varying light polarization \( \varrho \) and intensity \( I_L \). Here, we systematically adjust the physical parameters to study their effect on the scattering behavior.
In Figs. \ref{fig2}\textcolor{red}{(a-f)} with $\Delta = 0$  (no energy gap), distinct patterns of scattering resonances are observed. For $\varrho = 0$ in Figs. \ref{fig2}\textcolor{red}{(a-c)}, the scattering efficiency displays a series of bright regions that form well-defined structures as seen in Fig. \ref{fig2}\textcolor{red}{a}. In particular, below $R \approx 30$ nm, the interaction between the electron and the GQDs is minimal, a result consistent with previous research \cite{Azar24life,Penalight}. The resonances appear as curved bands that shift and intensify with increasing light intensity from $I_L = 1$ W/cm² to 5 W/cm², with the interaction starting at smaller values of $R$, as shown in Figs. \ref{fig2}\textcolor{red}{(b, c)}. When the polarization is increased to $\varrho = 0.5$ in Figs. \ref{fig2}\textcolor{red}{(d-f)}, the resonance patterns undergo significant changes, illustrating the profound effect of light polarization on the electron confinement mechanism.
Considering an energy gap of $\Delta = 20$ meV in Figs. \ref{fig2}\textcolor{red}{(g-l)}, the scattering phenomenon is significantly affected. The resonance patterns become more pronounced and show a significant shift in their positions compared to the scenario without a gap. This behavior is attributed to the modification of the electron's effective mass due to the energy gap, which changes the quantum confinement conditions. For $\varrho = 0$ in Figs. \ref{fig2}\textcolor{red}{(g-i)}, the resonances appear more sharply defined compared to those observed for $\Delta = 0$. The combination of high light intensity ($I_L = 5$ W/cm²) and the presence of a finite energy gap leads to the strongest modulation of the scattering efficiency.
Of particular note is the appearance of well-defined regions of high scattering efficiency, which become more pronounced in the presence of the energy gap. These localized regions of enhanced scattering suggest the formation of quasi-bound states within the quantum dot, where electron confinement is enhanced by the combined effects of the magnetic field and the laser-induced gap. These regions highlight how the interaction between the magnetic confinement and the energy gap creates favorable conditions for electron localization, resulting in enhanced scattering processes. The dependence on the radius $R$ shows periodic-like behavior, indicating size-dependent resonances characteristic of quantum confinement effects. In addition, the magnetic field dependence shows a non-monotonic trend, with optimal scattering occurring at specific field strengths that are influenced by both the GQD sizes and the laser parameters.

	\begin{figure}[ht]
	\centering
	\includegraphics[scale=0.32]{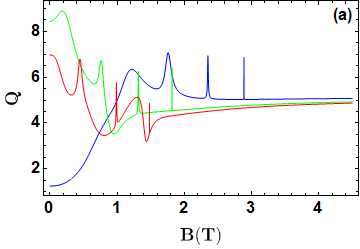}\includegraphics[scale=0.32]{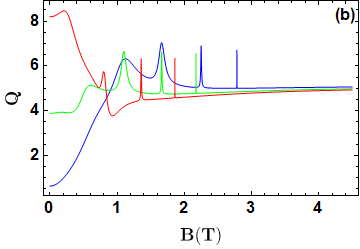}
	\includegraphics[scale=0.32]{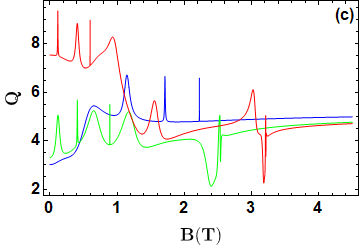}\includegraphics[scale=0.32]{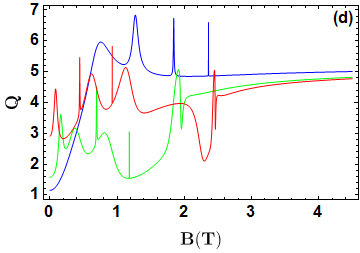}
	\caption{The scattering efficiency $Q$ as a function of the magnetic field strength $B$ for $E = 20 $ meV, $R = 70$ nm,  energy gap  [(a,b): $\Delta = 0 $, (c,d): $\Delta = 20 $ meV] and  light polarization  [(a,c): $\varrho = 0$, (b,d): $\varrho = 0.5$], and light intensity [blue line: $I_L= 1$ W/cm$^2$, green line: $I_L = 3$ W/cm$^2$, red line: $I_L = 5$ W/cm$^2$].}
	\label{fig3}
\end{figure}

Fig. \ref{fig3} shows the scattering efficiency $Q$ as a function of the magnetic field strength $B$ for a quantum dot radius of $R = 70$ nm and an incident electron energy of $E = 20$ meV. We systematically explore the effects of varying the energy gap $\Delta$, the light polarization $\varrho$, and the light intensity $I_L$, providing a detailed study of how these parameters influence the scattering behavior.
In Fig. \ref{fig3}\textcolor{red}{(a,b)} for $\Delta=0$, several key features emerge. For unpolarized light, Fig. \ref{fig3}\textcolor{red}{a} shows  that the scattering efficiency exhibits distinct resonance peaks whose amplitudes are strongly influenced by the light intensity. As the incident light intensity $I_L$ increases from 1 W/cm² (blue line) to 5 W/cm² (red line), a significant enhancement of the scattering peaks is observed, especially in the magnetic field region of $B = 2\sim3$ T. When the light polarization is increased to $\varrho = 0.5$ in Fig. \ref{fig3}\textcolor{red}{b}, the resonance structure changes, exhibiting shifts in peak positions and variations in their relative intensities.
The introduction of an energy gap $\Delta = 20$ meV in Fig. \ref{fig3}\textcolor{red}{(c,d)} has a significant effect on the scattering behavior. For $\varrho = 0$ in Fig. \ref{fig3}\textcolor{red}{c}, the resonance peaks become sharper and more pronounced compared to the case without gap. In addition, the peak positions shift systematically with increasing light intensity, indicating a strong interaction between the laser-induced gap and the magnetic confinement. In Fig. \ref{fig3}\textcolor{red}{d}, where $\varrho = 0.5$, the most dramatic changes in the scattering profile are observed, with enhanced resonance amplitudes and noticeable shifts in their positions.
A particularly intriguing aspect is the appearance of multiple resonance peaks that become more pronounced at higher light intensities, especially when the energy gap is present. These resonances are related to the formation of Landau-like levels, which are modified by the laser field, resulting in enhanced electron confinement at certain magnetic field strengths. The interaction between the magnetic field and the laser parameters (intensity and polarization) creates a dynamic platform for fine-tuning the electron scattering characteristics in the GQDs.

\begin{figure}[ht]
	\centering
	\includegraphics[scale=0.165]{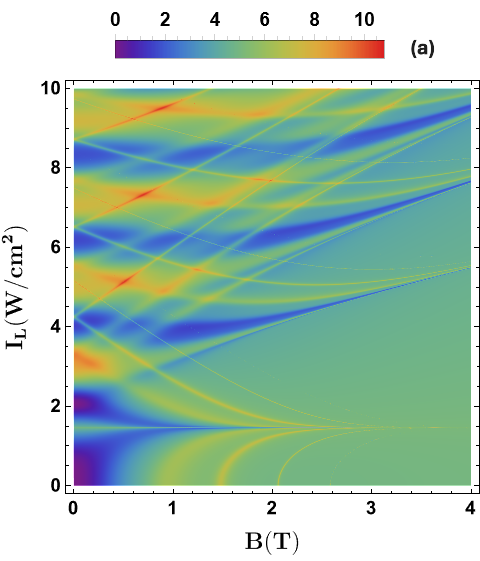}
	\includegraphics[scale=0.165]{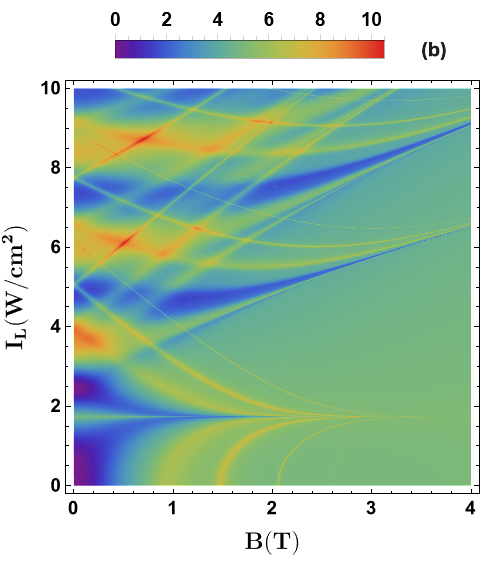}
	\includegraphics[scale=0.165]{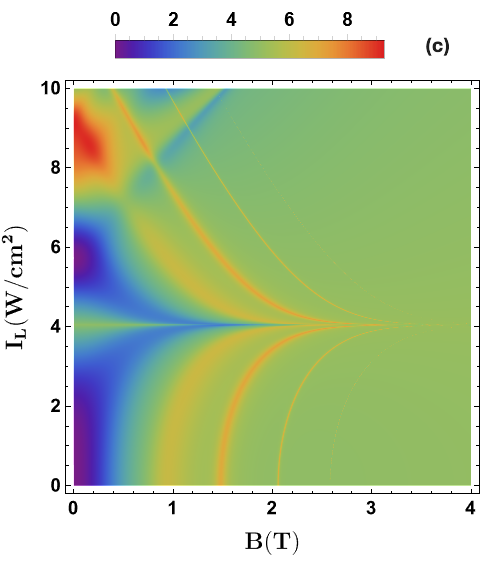}\\
	\includegraphics[scale=0.165]{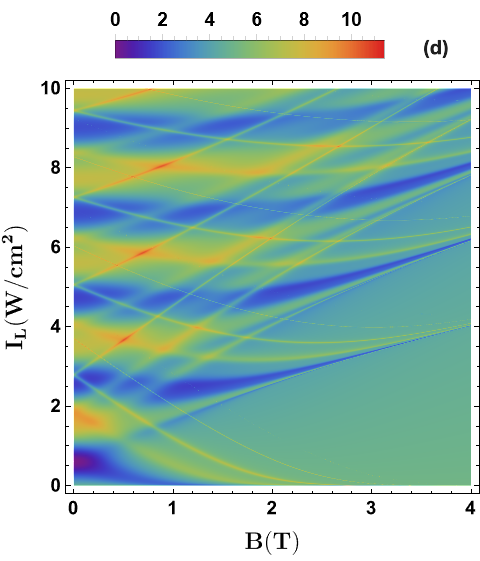}
	\includegraphics[scale=0.165]{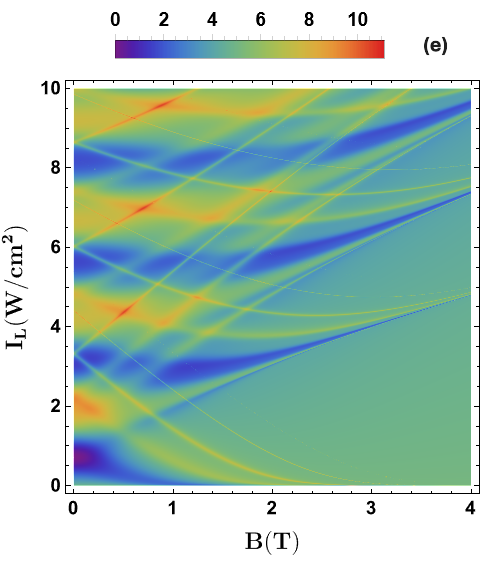}
	\includegraphics[scale=0.165]{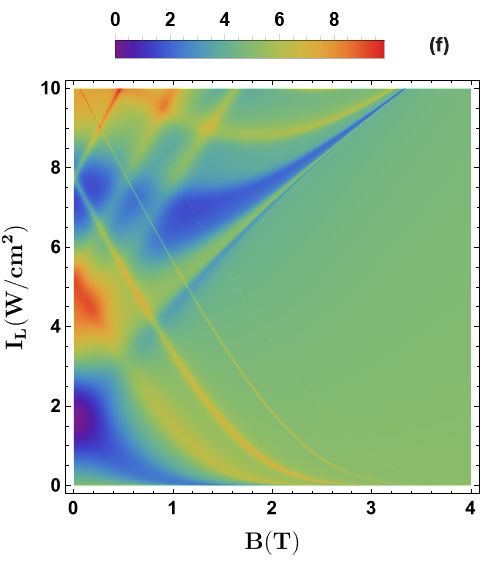}
	\caption{The scattering efficiency $Q$ as a function of the magnetic field strength $B$ and light intensity $I_L$ for $E = 20$ meV, $R = 70$ nm,  energy gap [(a,b,c): $\Delta = 0$ meV, (d,e,f): $\Delta = 20$ meV], and  light polarization [(a,d): $\varrho = 0$, (b,e): $\varrho = 0.4$, (c,f): $\varrho = 0.8$].}
	\label{fig4}
\end{figure}

Fig. \ref{fig4} provides a visualization of the scattering efficiency $Q$ as a function of both magnetic field strength $B$ and light intensity $I_L$ for  $E = 20$ meV and $R = 70$ nm. Here, we examine the combined effects of the energy gap $\Delta$ and light polarization $\varrho$ over six different panels. 
In Figs. \ref{fig4}\textcolor{red}{(a-c)} for $\Delta = 0$, the scattering pattern for unpolarized light (panel a) shows well-defined resonance bands that extend diagonally across the $B$-$I_L$ plane, suggesting a correlated dependence on both parameters. As the polarization increases to 0.4 in Fig. \ref{fig4}\textcolor{red}{b}, these resonance bands change, exhibiting variations in both intensity and position. At a high polarization of 0.8 in Fig. \ref{fig4}\textcolor{red}{c}, the scattering pattern shows more pronounced changes, with a greater contrast between regions of high and low scattering efficiency.
We observe that for $\Delta = 20$ meV in Figs. \ref{fig4}\textcolor{red}{(d-f)}, significant transformations in the scattering efficiency occur in different polarization regimes.
For $\varrho = 0$ in Fig. \ref{fig4}\textcolor{red}{d}, the resonance structure is more clearly defined compared to the case without a gap, showing distinct regions of enhanced scattering efficiency. At an intermediate polarization of $\varrho = 0.4$ in Fig. \ref{fig4}\textcolor{red}{e}, a complex interaction between the magnetic field and light intensity generates a rich pattern of resonances. In the highest polarization case, $\varrho = 0.8$ in Fig. \ref{fig4}\textcolor{red}{f}, the scattering pattern undergoes the most significant modification, with highly localized areas of enhanced scattering. 
A key feature is the appearance of "hotspots" of high scattering efficiency for certain combinations of $B$ and $I_L$, which are particularly noticeable in the finite energy gap cases. These hotspots indicate optimal conditions where the interplay between magnetic confinement and laser-induced effects enhances electron scattering. The locations and intensities of these hotspots are highly sensitive to light polarization, providing a potential means of fine-tuning the scattering properties of the system.

\begin{figure}[ht]
	\centering
	\includegraphics[scale=0.215]{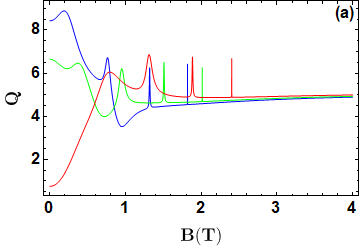}
	 \includegraphics[scale=0.225]{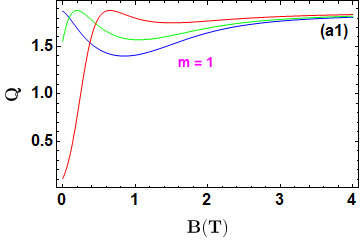} 
	 \includegraphics[scale=0.225]{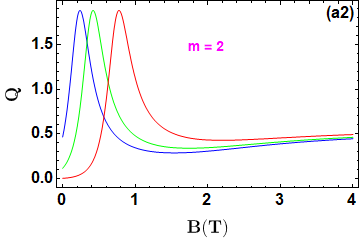}
	 \includegraphics[scale=0.225]{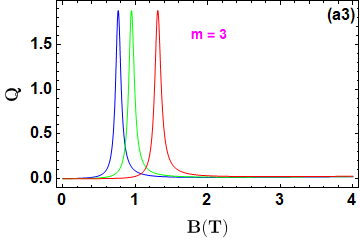} 
	 \includegraphics[scale=0.225]{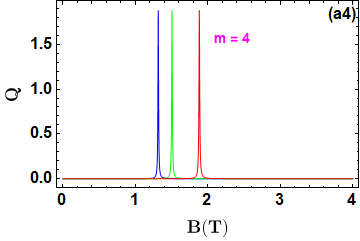}\\ 
	\includegraphics[scale=0.215]{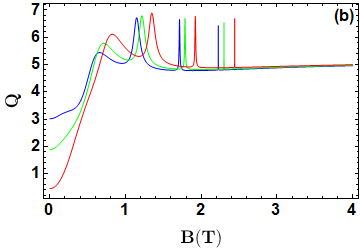}
	\includegraphics[scale=0.225]{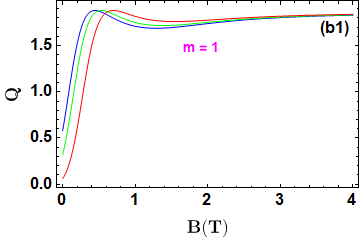} 
	\includegraphics[scale=0.225]{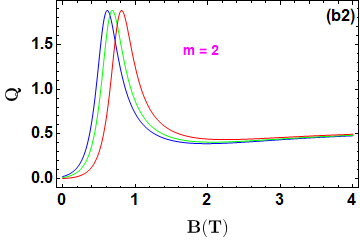}
	\includegraphics[scale=0.225]{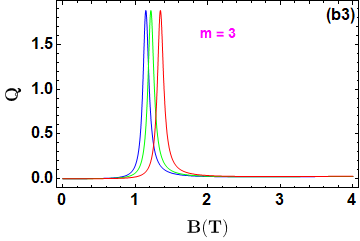} 
	\includegraphics[scale=0.225]{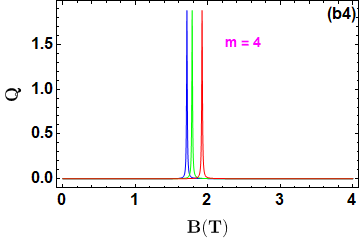}
	\caption{The scattering efficiency $Q$ as a function of the magnetic field strength $B$  for $E = 20$ meV, $R = 70$ nm,  energy gap and light intensity [(a,a1,a2,a3,a4): ($\Delta = 0$, $I_L = 3$ W/cm²),  (b,b1,b2,b3,b4): ($\Delta = 20$ meV, $I_L = 1$ W/cm²)], and light polarization [blue line: $\varrho = 0$, green line: $\varrho = 0.3$,  red line: $\varrho = 0.5$]. In panels (a1,a2,a3,a4) and (b1,b2,b3,b4), we present $Q$ separately for each excited mode, as indicated in the plots.}
	\label{fig5}
\end{figure}

Fig. \ref{fig5} shows the scattering efficiency $Q$ and its constituent excited modes as a function of the magnetic field strength $B$ for $E = 20$ meV and $R = 70$ nm. The analysis is divided into two main scenarios: one without a gap at moderate light intensity ($\Delta=0$, $I_L = 3$ W/cm²) and one with a gap at lower light intensity ($\Delta=20$ meV, $I_L = 1$ W/cm²).
For $\Delta=0$ in Figs. \ref{fig5}\textcolor{red}{(a, a1-a4)}, it is clear that the total scattering efficiency (panel a) results from the combination of four different excited modes shown in Figs. \ref{fig5}\textcolor{red}{(a1-a4)}. Each mode exhibits different resonance patterns and different responses to changes in light polarization. As the polarization increases from $\varrho=0$ (blue line) to $\varrho=0.5$ (red line), we observe systematic shifts in resonance positions and changes in peak intensities. In particular, the lower-order modes (Figs. \ref{fig5}\textcolor{red}{(a1, a2)}) exhibit more pronounced resonant properties compared to the higher-order modes (Figs. \ref{fig5}\textcolor{red}{(a3, a4)}), suggesting that these lower-order modes contribute more significantly to the overall scattering efficiency.
For $\Delta = 20$ meV in Figs. \ref{fig5}\textcolor{red}{(b, b1-b4)}, even at the lower light intensity of $I_L = 1$ W/cm², significant changes are observed in both the total scattering profile and the contributions of individual modes. The resonance peaks become sharper and more pronounced, indicating increased electron confinement. By separating the modes, we can observe how the energy gap affects different angular momentum channels differently, with certain modes showing greater sensitivity to polarization variations. This mode-specific analysis provides valuable insights into the quantum mechanical nature of the electron confinement and the scattering processes in the system.
Decomposing the total efficiency into its constituent modes reveals that the observed complex scattering behavior results from the intricate interplay between different angular momentum channels. Each channel responds differently to the combined effects of magnetic field, energy gap, and light polarization. A thorough understanding of how these modes contribute can be critical for applications that require the selective excitation of specific electron states.

\begin{figure}[ht]
	\centering
	\includegraphics[scale=0.17]{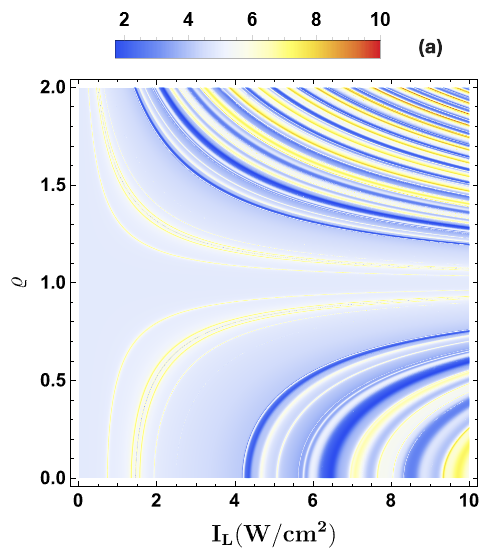}\includegraphics[scale=0.17]{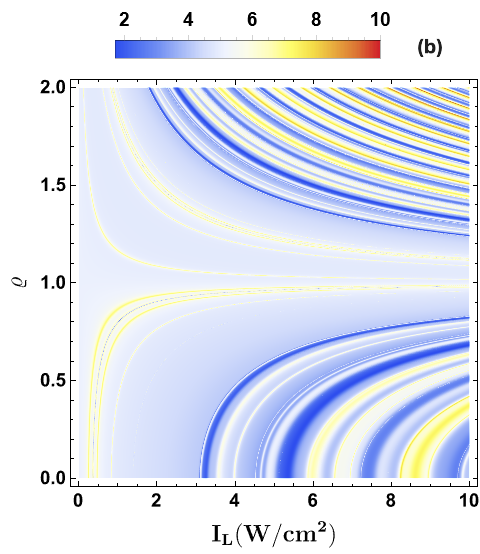}\includegraphics[scale=0.17]{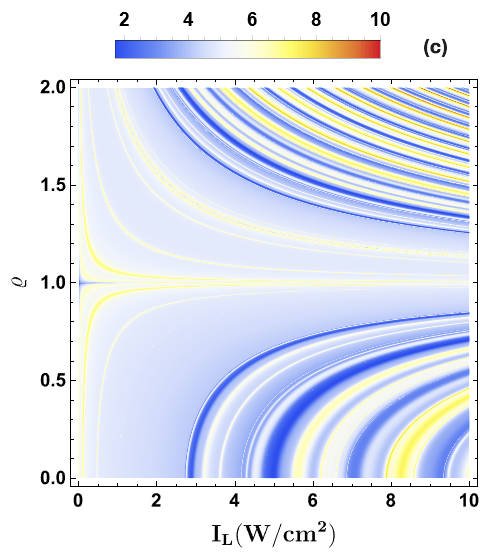}
	\caption{The scattering efficiency $Q$ as a function of the light intensity $I_L$ and light polarization $\varrho$ for $B = 2.2$ T, $E = 20$ meV, $R = 70$ nm, and  energy gap [(a):  $\Delta = 0$, (b): $\Delta = 15$ meV,  (c):  $\Delta = 20$ meV].}
	\label{fig6}
\end{figure}
In Fig. \ref{fig6}, we present the scattering efficiency $Q$ as a function of light intensity $I_L$ and polarization $\varrho$. This analysis is performed under the following specific conditions: magnetic field strength $B = 2.2$ T, electron energy $E = 20$ meV, and quantum dot radius $R = 70$ nm. Fig. \ref{fig6} examines three different energy gap regimes, providing valuable insight into how the energy gap affects the scattering properties of the system.
For $\Delta=0$ in Fig. \ref{fig6}\textcolor{red}{a}, the scattering pattern shows a clear dependence on both light intensity and polarization, with noticeable variations across these parameters. The efficiency curve shows alternating regions of high and low scattering, creating characteristic patterns. However, when an intermediate energy gap is introduced in Fig. \ref{fig6}\textcolor{red}{b}, these patterns undergo a significant change, accompanied by an increased contrast between regions of maximum and minimum scattering. The scattering efficiency displays more pronounced variations in both light intensity and polarization, indicating enhanced electron confinement effects.
As shown in Fig. \ref{fig6}\textcolor{red}{c}, at the highest energy gap ($\Delta = 20$ meV) the scattering pattern undergoes a significant transformation, building on the trends observed in Figs. \ref{fig6}\textcolor{red}{(a,b)}. The regions of increased scattering efficiency become more sharply defined and localized to specific regions. This increased localization marks a natural progression from the broader patterns in Fig. \ref{fig6}\textcolor{red}{a} and the intermediate structure in Fig. \ref{fig6}\textcolor{red}{b}. Notably, the maximum scattering efficiency occurs at certain combinations of light intensity and polarization, indicating optimal conditions for electron confinement. The emergence of these well-defined regions of high scattering efficiency highlights how the larger energy gap increases the sensitivity of the system to both light intensity and polarization. This increased sensitivity suggests a stronger coupling between electron dynamics and laser field parameters, providing a more precise mechanism for controlling electron transport within the quantum dot structure.

\begin{figure}[ht]
 	\centering
 	\includegraphics[scale=0.32]{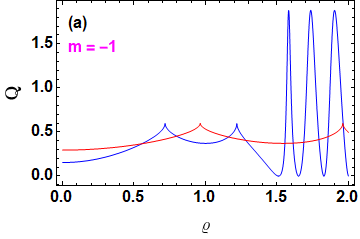} \includegraphics[scale=0.32]{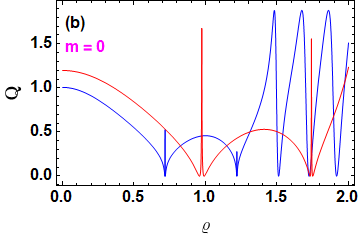} \includegraphics[scale=0.32]{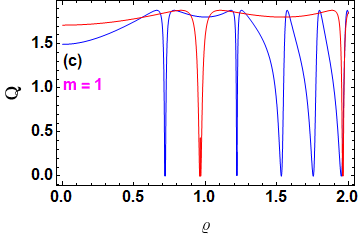}
 	\includegraphics[scale=0.32]{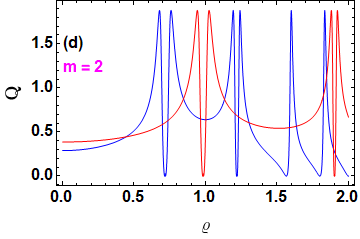} 
 	\includegraphics[scale=0.32]{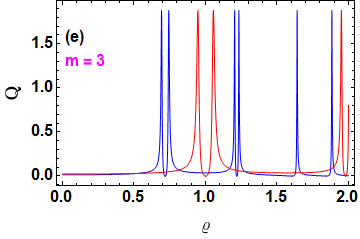}\includegraphics[scale=0.32]{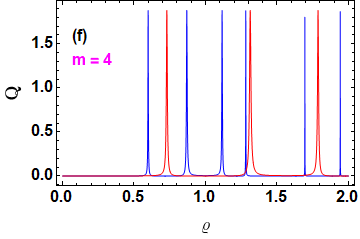}
 	\caption{The scattering efficiency $Q$ as a function of the light polarization $\varrho$ for $E = 20$ meV, $R = 70$ nm,   energy gap and light intensity [blue line: ($\Delta = 0$, $I_L = 3$ W/cm²), red line: ($\Delta = 20$ meV, $I_L = 1$ W/cm²)]. In panels (a,b,c,d,e,f), we present $Q$ for each excited mode individually, as indicated in the plots.}
 	\label{fig7}
 \end{figure}

As shown in Fig. \ref{fig7}, a mode-by-mode analysis of the scattering efficiency $Q$ with light polarization $\varrho$ is presented for two configurations: the no-gap case with $I_L = 3$ W/cm² (blue line) and the finite-gap case with $\Delta = 20$ meV and $I_L = 1$ W/cm² (red line). Each mode is analyzed separately.
For the diffusion mode $m=-1$ (Fig. \ref{fig7}\textcolor{red}{a}), the no-gap case shows a consistent decrease in scattering efficiency with increasing polarization. However, when an energy gap is introduced, the mode exhibits enhanced scattering efficiency with more pronounced fluctuations. This suggests that the gap significantly affects the confinement of electrons in negative angular momentum states.
In Fig. \ref{fig7}\textcolor{red}{b}, for the fundamental mode ($m=0$), the behavior is different in both configurations. In the no-gap case, the scattering efficiency remains relatively constant over a range of polarization values. In the finite-gap case, however, it shows more pronounced variations with well-defined maxima and minima, indicating a stronger coupling between the electron states and the laser field.
For the diffusion mode $m=1$ (Fig. \ref{fig7}\textcolor{red}{c}), this positive angular momentum mode shows more structured variations in both configurations. The finite gap configuration shows more pronounced modulation, suggesting that the first positive angular momentum state is particularly sensitive to the combined effects of gap and polarization. In Fig. \ref{fig7}\textcolor{red}{d}, the second order mode ($m=2$) shows a complex structure with multiple features. The no-gap case shows moderate oscillations, while the finite gap case shows more dramatic variations with polarization, indicating a significant change in electron confinement at this angular momentum.For the diffusion mode $m=3$ (Fig. \ref{fig7}\textcolor{red}{e}), this mode shows a decrease in overall amplitude compared to the lower-order modes, but still retains distinct features between the two configurations. The finite-gap case shows more structured variations with polarization, highlighting the persistence of quantum effects at higher angular momenta.In Fig. \ref{fig7}\textcolor{red}{f}, the highest-order mode ($m=4$) shows the smallest amplitude of all modes, but still retains distinguishable features between the no-gap and finite-gap cases. Even at this higher angular momentum state, the energy gap continues to influence the scattering behavior, albeit with reduced overall efficiency.

\subsection{Lifetime time and density}

To analyze the transient nature of the quasi-bound states in our system, we investigate their lifetime (trapping time). Unlike classical bound states, which are characterized by a negative discrete energy spectrum, quasi-bound states have positive energy in the continuum. A technique based on the complex energy of the incident electron is required to study such a time \cite{Narimanov99}. Indeed, we can decompose the incident energy as
\begin{equation}
	E = E_r - iE_i 
	\end{equation}
where $E_r$ is the resonance energy (real part) and $E_i > 0$ is related to the lifetime $\tau$ by
\begin{equation}
	\tau = \frac{\hbar}{E_i}.
	\end{equation}
	Since we have $E = v_F\hbar k$, then we can write 
 the wave vector as a complex number
\begin{equation}
	k = k_r - ik_i
	\end{equation}
and therefore we establish the relation
\begin{equation}
	\tau = \frac{1}{v_F k_i}.
	\end{equation}
Each diffusion mode \( m = \dots, -2, -1, 0, 1, 2, \dots \) has different characteristics in terms of diffusion efficiency and density. To determine the lifetime corresponding to each mode, we impose the following boundary condition at the interface \( r = R \)
{\begin{equation}
	\frac{\psi^+_{\pm}(r)}{\psi^-_{\pm}(r)} = \left.\frac{H_m(kr)}{H_{m+1}(kr)}\right|_{r=R}.
	\end{equation}

Fig. \ref{fig8} shows the lifetime $\tau$ of quasi-bound states as a function of magnetic field strength $B$ for different configurations. The analysis is performed for different modes under two conditions: ($\Delta = 0$ meV, $I_L = 1$ W/cm²) in Figs. \ref{fig8}\textcolor{red}{(a,c,e)} and ($\Delta = 20$ meV, $I_L = 3$ W/cm²) in Figs. \ref{fig8}\textcolor{red}{(b,d,f)}, with different light polarization values ($\varrho = 0, 0.3, 0.5$).As seen in Fig. \ref{fig8}, there is a clear trend of increasing trapping time $\tau$ with increasing magnetic field strength $B$. A notable observation is the difference in behavior between modes: the $m = 2$ mode (Fig. \ref{fig8}\textcolor{red}{(a,b)}) shows significant trapping effects at lower magnetic field strengths, while higher modes require stronger fields to exhibit noticeable trapping times. This difference is particularly evident in both the gapless and finite gap configurations.
In the gapless configuration (Fig. \ref{fig8}\textcolor{red}{(a,c,e)}) with $I_L = 1$ W/cm², the trapping time varies significantly with light polarization. Starting from a magnetic field value of about $B\sim2$ T, $\tau$ increases progressively with $\varrho$. At $B = 4.5$ T, the lifetimes are $\tau = 2.2$ ps at $\varrho = 0$, $\tau = 2.5$ ps at $\varrho = 0.3$, and $\tau = 4.2$ ps at $\varrho = 0.5$.
When an energy gap is introduced (Fig. \ref{fig8}\textcolor{red}{(b,d,f)}), the trapping times are significantly increased. This increase in trapping time with magnetic field demonstrates the strong effect of magnetic confinement in the system. The results are in agreement with literature observations and provide new insights into the combined role of light polarization and energy gap in influencing the trapping dynamics of electrons.

\begin{figure}[ht]
	\centering
	\includegraphics[scale=0.22]{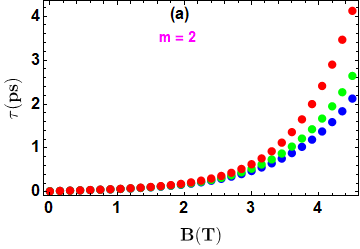} \includegraphics[scale=0.22]{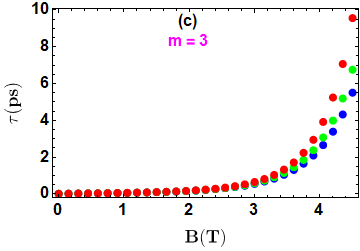} \includegraphics[scale=0.22]{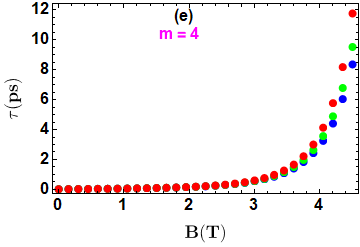}\\
	\includegraphics[scale=0.22]{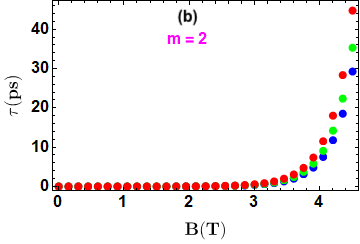} \includegraphics[scale=0.22]{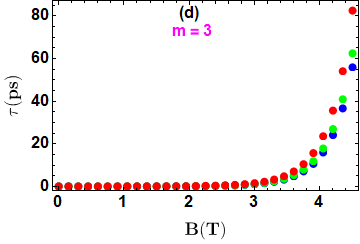} \includegraphics[scale=0.22]{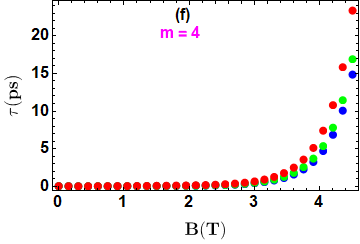}
	\caption{The life time  $\tau$ as a function of the magnetic field strength $B$  for $E = 20$ meV, $R = 70$ nm, light polarization [blue line: $\varrho = 0$, green line: $\varrho = 0.3$, red line: $\varrho = 0.5$], energy gap and light intensity  [(a,c,e): $(\Delta = 0$, $I_L = 1$ W/cm$^2$), (b,d,f): $(\Delta = 20$ meV, $I_L = 3$ W/cm$^2$)]. We present the life time separately for each mode, as indicated in the plots. }
	\label{fig8}
\end{figure}

 \begin{figure}[ht]
	\centering
	\includegraphics[scale=0.22]{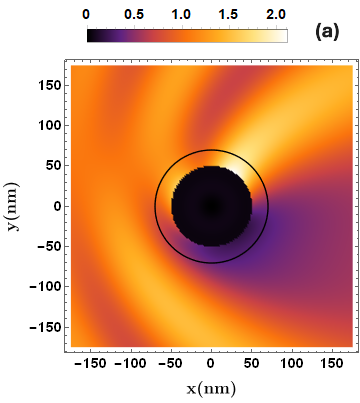} \includegraphics[scale=0.22]{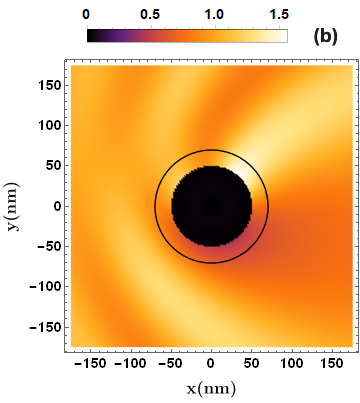} \includegraphics[scale=0.22]{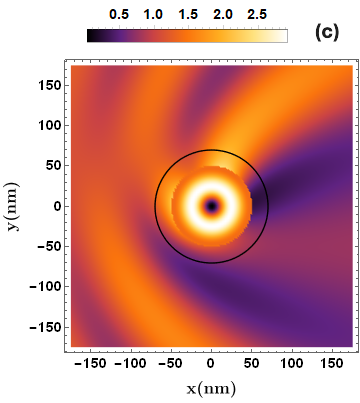}
	\caption{Density $\rho=\Phi^\dagger \Phi$ for a real space with  $E=20$ meV, $\varrho = 0$, $R=70$ nm and the parameters [(a): ($B=1.47$ T, $\Delta = 0$, $I_L = 3$ W/cm2), (b): ($B = 1.45$ T, $\Delta = 10$ meV, $I_L = 3$ W/cm2),  (c):  ($B = 2.04$ T, $\Delta = 20$ meV, $I_L = 1$ W/cm2)]. The spatial localization of the GQDs is marked by black circles in the density plots.}
	\label{fig9}
\end{figure}

As shown in Fig. \ref{fig9}, an investigation of electron scattering in the GQDs under the combined influence of a magnetic field $B$, an energy gap $\Delta$, and circularly polarized laser irradiation is presented. This density study ($\rho=\Phi^\dagger \Phi$) explores how these parameters affect electron confinement, addressing a significant challenge posed by the Klein tunneling effect inherent in graphene.
In the initial configuration (Fig. \ref{fig9}\textcolor{red}{a}: $B = 1.47$ T, $\Delta = 0$ meV, $I_L = 3$ W/cm²), the electron density shows concentric diffraction patterns around the graphene quantum dot (GQD) with reduced central intensity. This asymmetric spiral shape clearly illustrates the Klein tunneling effect, where electrons escape from the quantum dot by diffraction instead of being confined. In Fig. \ref{fig9}\textcolor{red}{b}, the introduction of a moderate energy gap ($B = 1.45$ T, $\Delta = 10$ meV, $I_L = 3$ W/cm²) significantly changes the electron distribution, resulting in an increased density at the center of the GQD, while the diffraction patterns persist. This intermediate configuration suggests the formation of quasi-bound states, demonstrating the beneficial effect of the combination of magnetic field and energy gap on electron confinement.
The optimal configuration (Fig. \ref{fig9}\textcolor{red}{c}: $B = 2.04$ T, $\Delta = 20$ meV, $I_L = 1$ W/cm²) shows exceptional electron confinement. The increased energy gap, combined with a reduction in light intensity, results in a significant concentration of density in the center of the GQDs. The clear circular symmetry of the distribution and the near absence of diffraction patterns indicate efficient and stable electron confinement. These results demonstrate that a well-designed combination of magnetic field, energy gap, and laser irradiation can effectively mitigate the Klein tunneling effect and facilitate electron confinement in the GQDs. In particular, the observation that reduced light intensity along with an enhanced energy gap promotes electron trapping offers promising ways to control electronic states in graphene-based devices.

	\section{Conclusion} \label{cc}
	
	In the present study, a comprehensive theoretical investigation of the electronic interactions in the graphene quantum dots (GQDs) under the combined influence of a uniform magnetic field, an energy gap, and circularly polarized laser irradiation has been carried out. The approach, based on the solution of the Dirac equation with appropriate boundary conditions, led to the derivation of analytical expressions for key physical quantities, such as scattering efficiency, electron densities, and quasi-bound state lifetimes. The numerical results demonstrate controlled electron confinement in the GQDs due to the synergistic interaction of these external parameters.
	Our results reveal several significant phenomena. In the absence of an energy gap, electrons undergo diffraction within the the GQDs with minimal interaction, resulting in a reduced electron density at the center. However, when a magnetic field and an energy gap are introduced along with laser irradiation, the concentration of electrons in the GQDs increases. This results in higher trapping probabilities, longer quasi-bound state lifetimes, and improved scattering efficiency.
	Furthermore, the efficiency of electron trapping can be precisely controlled by adjusting the light intensity, polarization, and energy gap amplitude. In particular, the energy gap plays a critical role in influencing the spatial localization of the electrons and the lifetimes of the quasi-bound states, providing a powerful mechanism for tuning the electronic properties of graphene quantum dots.
	
	Our results underscore the significant potential for manipulating electron confinement in graphene quantum dots through precise control of external parameters such as magnetic fields, energy gaps, and laser irradiation. By systematically investigating these factors, this work provides critical insights into the mechanisms that govern electron trapping and scattering efficiency in idealized GQD systems. Although the study is based on simplified assumptions, it lays a solid foundation for future investigations that could incorporate additional complexities, such as electron-electron interactions, disorder effects, and the interplay between multiple quantum states, to reflect more realistic conditions.
	These results offer promising avenues for future research, particularly in the design of advanced graphene-based devices, where the precise tuning of electron confinement and transport properties is critical. By further exploring how these external parameters can be optimized, researchers could unlock new strategies for controlling electronic, optical and spintronic properties at the nanoscale. This could lead to the development of cutting-edge technologies in areas such as optoelectronics, spintronics, and quantum computing. In addition, this study highlights the need for a more comprehensive understanding of the complex interactions within graphene quantum dots under realistic conditions. Future work that integrates more detailed models-accounting for factors such as many-body effects, disorder, temperature variations, and interactions with surrounding materials-will be essential for translating these theoretical insights into practical, functional devices. In this regard, the current study serves not only as a stepping stone for the development of graphene-based applications, but also as a catalyst for advancing the broader field of quantum dot research.

\end{document}